\documentclass[prb,twocolumn,showpacs,superscriptaddress,floatfix, nofootinbib]{revtex4-2}
\usepackage{amsmath,amssymb,amsfonts,mathrsfs, amsbsy}
\usepackage{graphicx}
\usepackage[bookmarks=true, colorlinks=true, linkcolor=blue, urlcolor=blue, citecolor=blue, bookmarks=true, hyperindex=true]{hyperref}
\usepackage[normalem]{ulem}

\newcommand{\be}{\begin{equation}}
\newcommand{\ee}{\end{equation}}

\newcommand{\beq}{\begin{eqnarray}}
\newcommand{\eeq}{\end{eqnarray}}

\def\H1{\widehat{H}_1}

\newcommand{\ket}[1]{\left| #1 \right>}
\newcommand{\bra}[1]{\left< #1 \right|}

\begin{document}

\title{Adiabatic Transformations in Dissipative and Non-Hermitian Phase Transitions}

\author{Pavel Orlov}
\affiliation{Physics Department, Faculty of Mathematics and Physics, University of Ljubljana,
Ljubljana, Slovenia}
\affiliation{Russian Quantum Center, Skolkovo, Moscow 143025, Russia}
\affiliation{Nanocenter CENN, Jamova 39, SI-1000 Ljubljana, Slovenia}

\author{Georgy\,V.~Shlyapnikov}
\affiliation{Russian Quantum Center, Skolkovo, Moscow 143025, Russia}
\affiliation{Moscow Institute of Physics and Technology, Dolgoprudny, Moscow Region, 141701, Russia}
\affiliation{Université Paris-Saclay, CNRS, LPTMS, 91405 Orsay, France}
\affiliation{Van der Waals–Zeeman Institute, Institute of Physics, University of Amsterdam, Science Park 904, 1098 XH Amsterdam, The Netherlands}

\author{Denis\,V.~Kurlov}
\affiliation{Russian Quantum Center, Skolkovo, Moscow 143025, Russia}
\affiliation{National University of Science and Technology ``MISIS'', Moscow 119049, Russia}


\begin{abstract}
The quantum geometric tensor has established itself as a general framework for the analysis and detection of equilibrium phase transitions in isolated quantum systems. We propose a novel generalization of the quantum geometric tensor, which offers a universal approach to studying phase transitions in non-Hermitian quantum systems. 
Our generalization is based on the concept of the generator of adiabatic transformations and can be applied to systems described by either a Liouvillian superoperator or by an effective non-Hermitian Hamiltonian. We illustrate the proposed method by analyzing the non-Hermitian Su-Schrieffer-Heeger model and a generic quasi-free dissipative fermionic system with a quadratic Liouvillian.
Our findings reveal that this method effectively identifies phase transitions across all examined models, providing a universal tool for investigating general non-Hermitian systems.
\end{abstract}

\maketitle

\textit{\textbf{Introduction.}} ---
Equilibrium quantum phase transitions (QPTs) have been the subject of extensive research over recent decades~\cite{stanley1987introduction,Sachdev_2011}. One of the approaches to equilibrium QPTs in isolated quantum systems is based on information geometry and relies on the notion of the quantum geometric tensor (QGT), which was introduced in Ref.~\cite{Provost}.
The real part of the QGT coincides with the Fubini-Study metric tensor (also known as quantum metric), while the imaginary part is nothing but the Berry curvature. 
The QGT-based approach is particularly powerful because it does not require knowledge of an order parameter and can be applied universally to all equilibrium QPTs, including the topological ones. 
Indeed, a QPT is associated with the non-analytic change in the ground state of a system. Hence, phase transitions can be detected by the non-analytic behavior of the metric tensor~\cite{Zanardi_2006,zanardi2007differential,Campos2007} or the Berry curvature \cite{Thoules1984,bernevig2013topological,Carollo_2005,hamma2006berry}, i.e., through singular behavior of the QGT. 

In recent years, there has been significant interest in dissipative phase transitions, which occur in open quantum systems~\cite{Prosen_2008, Diehl_2008, Dalla_Torre_2010, Dagvadorj, Jin2016, Bakker2022, Kazmina2024}.
Partly, this interest is driven by remarkable experimental advances that enable the engineering and control of dissipation across various quantum systems, including ultracold neutral atoms~\cite{Mueller2012}, trapped ions~\cite{Leibfried2003}, quantum optical systems~\cite{Miri2019}, and superconducting circuits~\cite{Clerk2020}. 
These developments open up avenues for exploring the complex physics of dissipative and non-Hermitian (NH) quantum systems
, alongside potential applications in quantum technologies~\cite{PARIS2009, Harrington2022}.

For Markovian open systems described by a Liouvillian superoperator, phase transitions manifest themselves as a non-analytic behavior of the system's non-equilibrium steady state (NESS), which is generically a mixed state. Remarkably, the information-geometric approach to QPT can also be extended to dissipative phase transitions and mixed states.
Specifically, the Bures metric~\cite{Bures1969AnEO,UHLMANN1976273}, the Uhlmann curvature~\cite{UHLMANNphase}, and the quantum Fisher tensor~\cite{Ercolessi_2012, CarolloGEOMofQPT} generalize the Fubini-Study metric tensor, the Berry curvature, and the QGT, correspondingly.
Both the Bures metric and Uhlmann curvature have been successfully exploited to describe QPTs in open quantum systems~\cite{Zanardi2007BuresMO, Huang2014, Viyuela2014, Viyuela2014(2), Mera, Banchi_2014,carollo2017uhlmann}.
However, these quantum geometric measures for mixed states do not allow one to study phase transitions in quantum systems with non-Hermitian Hamiltonians. The latters appear as an effective phenomenological description of an open system and can be useful for the investigation of a wide range of physical phenomena, such as NH topological~\cite{Gong_2018} and parity-time (${\cal PT}$) symmetric phases~\cite{ElGanainy2018NonHermitianPA}. 
Various generalizations of the QGT have been proposed to deal with NH Hamiltonians~\cite{Brody_2013,Zhang_2019,Zhu_2021,ye2023quantum}. However, all proposed measures are either not gauge invariant, or vanish identically for kinematic reasons when applied to Liouvillian superoperators.
Thus, at present there are two distinct sets of measures for open and NH quantum systems, since no existing quantity applies to the study of phase transitions in both Liouvillians and NH Hamiltonians.

In this work we introduce a novel non-Hermitian generalization of the quantum geometric tensor (NH-QGT). It is gauge invariant and can be used for generic non-Hermitian systems, including those described by Liouvillians and NH Hamiltonians. Thus, the proposed generalization provides a universal framework for investigating phase transitions in such systems. 
Our generalization is based on the generator of adiabatic transformations, known as the adiabatic gauge potential (AGP), which has proven to be extremely useful for isolated systems~\cite{Kolodrubetz_2017, Pandey_2020}.
We test our proposal on a few exactly solvable models with a rich phase diagram structure, where the NH-QGT can be calculated explicitly. The newly proposed NH-QGT correctly captures critical points in all analyzed cases.

\textit{\textbf{AGP and NH-QGT.}}
---
For an isolated system described by a  parameter-dependent Hermitian Hamiltonian $\hat{H}(\boldsymbol{\lambda})$, the QGT for an eigenstate $\ket{n(\boldsymbol{\lambda)}}$ is defined as \cite{Provost}
\begin{equation}\label{QGT_pure}
    \chi^{(n)}_{\mu \nu} =  \langle \partial_{\mu} n | \partial_{\nu} n  \rangle - \langle \partial_{\mu} n | n  \rangle \langle n  | \partial_{\nu} n  \rangle. 
\end{equation}
There are several noteworthy properties associated with this quantity. To begin with, QGT is both Hermitian and positive-semidefinite. Its real part, denoted as $g_{\mu \nu}^{(n)} = \text{Re}[ \chi_{\mu \nu}^{(n)}]$, is a symmetric positive-semidefinite tensor known as the Fubini-Study metric, while its imaginary part, $F_{\mu \nu}^{(n)} = - \frac{1}{2} \text{Im} [ \chi_{\mu \nu}^{(n)} ]$, is an anti-symmetric tensor known as the Berry curvature \cite{Kolodrubetz_2017}. Additionally, QGT exhibits gauge invariance under the transformation $\ket{n} \rightarrow e^{i \gamma (\boldsymbol{\lambda})} \ket{n}$, related to the phase ambiguity of eigenstates. The choice of the second term in Eq.~(\ref{QGT_pure}) is specifically made to ensure the gauge invariance.

Let us now consider a system described by a non-Hermitian matrix $K$, which can be either a non-Hermitian effective Hamiltonian $H$ or a Liouvillean superoperator $\mathcal{L}$. In the latter case, we treat the matrix space as a vector space with the inner product defined as $\langle \rho_1 | \rho_2 \rangle = \text{Tr}(\rho_1^{\dagger} \rho_2)$, consequently considering the superoperator as an ordinary (albeit non-Hermitian) operator within this Hilbert space.

We now briefly summarize some key properties of non-Hermitian operators that are important for our purposes (see Ref.~\cite{Ashida2020} for a detailed overview).
For non-Hermitian operators the right and left eigenvectors do not coincide and one has
\begin{equation}
    K \ket{n_R} = \Lambda_n \ket{n_R}, \quad K^{\dagger} \ket{n_L} = \Lambda_n^{*} \ket{ n_L }.
\end{equation}
Moreover, the left and right eigenvectors are not orthogonal individually, e.g. $\langle m_R | n_R \rangle = C_{mn} \neq \delta_{mn}$, but together they form a biorthogonal system, satisfying the relation $\langle m_L | n_R \rangle = \delta_{mn}$. The Gram matrices for the left and right eigenvectors are inverses of each other, i.e. $\langle m_L | n_L \rangle = (C^{-1})_{mn}$.

For non-Hermitian systems, eigenstates are defined not merely up to a phase but up to an arbitrary non-zero constant. This implies that any generalization of QGT (\ref{QGT_pure}) must remain invariant under gauge transformations 
\begin{equation} \label{gauge_transform_NH_eigenstates}
    \ket{n_R} \rightarrow e^{r_n(\boldsymbol{\lambda})} \ket{n_R }, \qquad \ket{n_L} \rightarrow e^{-r_n^*(\boldsymbol{\lambda})} \ket{n_L},     
\end{equation}
where $r_n(\boldsymbol{\lambda})$ is an arbitrary complex-valued function. 


To construct a non-Hermitian generalization of the quantum geometric tensor, we introduce the generator of adiabatic transformations $\mathcal{A}_{\lambda} \ket{ n_R}= \ket{\partial_{\lambda} n_R}$, known as the adiabatic gauge potential (AGP) \cite{AGPnote, Kolodrubetz_2017}.
From the biorthogonality condition one has 
$\mathcal{A}^{\dagger}_{\lambda} \ket{ n_L} = - \ket{\partial_{\lambda} n_L}$. For Hermitian systems, the left and right eigenvectors coincide, so that ${\cal A}_{\lambda}$ is anti-Hermitian.  As a result, the Hermitian QGT (\ref{QGT_pure}) can be expressed as a connected part of the expectation value in two equivalent ways:
\begin{equation} \label{QGT_AGP_2_ways}
    \chi_{\mu \nu}^{(n)} = - \langle \mathcal{A}_{\mu} \mathcal{A}_{\nu} \rangle_c  = \langle \mathcal{A}_{\mu}^{\dagger} \mathcal{A}_{\nu} \rangle_c.
\end{equation}
However, for non-Hermitian systems, the AGP is no longer anti-Hermitian.
Thus, Eq.~(\ref{QGT_AGP_2_ways}) suggests that a non-Hermitian generalization of the QGT~$\chi_{\mu \nu}^{(n)}$ can be constructed in two ways. 

One approach uses $-\bra{n_L} \mathcal{A}_{\mu} \mathcal{A}_{\nu} \ket{ n_R }$ as a basis and then includes necessary counter-terms to ensure gauge invariance. In this way, 
 the first generalization is given by
\begin{equation}\label{LR-QGT}
    \eta_{\mu \nu}^{(n)}  =  \langle \partial_{\mu} n_L | \partial_{\nu} n_R  \rangle - \langle \partial_{\mu} n_L | n_R  \rangle \langle n_L  | \partial_{\nu} n_R  \rangle.
\end{equation}
This tensor has been already studied in a series of works~\cite{Brody_2013,Zhang_2019,Zhu_2021,ye2023quantum} in the context of criticality in non-Hermitian Hamiltonians. 
Despite its successful application in describing phase transitions, this quantity cannot be utilized as a probe of criticality in NESS-QPTs. 
This is attributed to the fact that the left eigenvector, which corresponds to NESS, is the identity operator \cite{Identity-left-eigen}. Consequently, for NESS $\eta_{\mu \nu}$ is trivially equal to zero.

This circumstance motivates us to define the second generalization of QGT. The same methodology can be applied if we use $ \bra{n_L} \mathcal{A}_{\mu}^{\dagger} \mathcal{A}_{\nu} \ket{ n_R }$ as the starting point. 
The relation $\bra{n_L} \mathcal{A}_{\mu}^{\dagger} = \sum_{m} (C^{-1})_{nm} \bra{\partial_{\mu} m_R}$ allows us to readily introduce the required counter-terms. Ultimately, we define our NH-QGT as

\begin{equation}\label{Gen-QGT}
    \zeta_{\mu \nu}^{(n)} = \sum_{m} (C^{-1})_{nm} \bra{ \partial_{\mu} m_R } \hat{1} - P_{m}^{\dagger} - P_{n} + P_{m}^{\dagger} P_n \ket{\partial_{\nu} n_R},
\end{equation}
where $P_{m} = \ket{m_R} \bra{m_L}$ represents a set of non-orthogonal projectors on right eigenstates.

The definition~(\ref{Gen-QGT}) can be expressed in a simpler form. Gauge transformations of eigenstates allow us to nullify the diagonal matrix elements of AGP (for a more detailed discussion of AGP properties, see the SM). In this gauge, the NH-QGT can be written as
\begin{equation}\label{Gen-QGT-AGP}
\begin{aligned}
    \zeta_{\mu \nu}^{(n)} = \bra{n_L} \mathcal{A_{\mu}^{\dagger} \mathcal{A}_{\nu}} \ket{ n_R }, 
\end{aligned}
\end{equation}
with $\mathcal{A}_{\lambda} $ such that $\forall m \text{ }\bra{m_L} \mathcal{A}_{\lambda} \ket{m_R} = 0$. We will use this expression below to calculate geometric tensor in dissipative models. Also, note that this definition can be straightforwardly generalized to the case of non-diagonalizable $K$ as well~\cite{non-diag}.

In general, the NH-QGT (\ref{Gen-QGT}) is neither Hermitian nor positive-semidefinite. However, for NESSes its complex structure simplifies. 
In this case the NH-QGT (\ref{Gen-QGT}) is purely real since for Liouvillean systems AGP operators preserve hermiticity and trace. Another important point is that the NH-QGT is not related to the geometry of a particular state, as it includes derivatives and the Gram matrix with the other eigenstates. 

To motivate the introduction of $\zeta_{\mu \nu}^{(n)}$ even more, we define the covariant derivative of eigenstates as the gauge covariant part of the derivative operator
\begin{equation}
\begin{aligned}
    &\ket{D_{\mu} n_R} \equiv \ket{\partial_{\mu} n_R} - A_{\mu}^{(n)} \ket{n_R}, \\
    &\bra{D_{\mu} n_L} \equiv \bra{\partial_{\mu} n_L} + A_{\mu}^{(n)} \bra{n_L},
\end{aligned}
\end{equation}
where $A_{\mu}^{(n)} = \langle n_L | \partial_{\mu} n_R \rangle = - \langle \partial_{\mu} n_L | n_R \rangle $ is a non-Hermitian analog of the Berry connection. Under the gauge transformation~(\ref{gauge_transform_NH_eigenstates}) 
the covariant derivatives are transformed simply as
\begin{equation} \label{cov_derivatives_transform}
\begin{aligned}
    &\ket{D_{\mu} n_R} \rightarrow e^{r_n(\boldsymbol{\lambda})} \ket{D_{\mu} n_R}, \\
    &\bra{D_{\mu} n_L} \rightarrow e^{-r_n(\boldsymbol{\lambda})} \bra{D_{\mu} n_L}.
\end{aligned}
\end{equation}
Using these derivatives, Eq.~(\ref{Gen-QGT}) can be rewritten in a form with a more transparent gauge invariance:

\begin{align}
    &\zeta_{\mu \nu}^{(n)} = \sum_{m} \langle n_L | m_L \rangle \langle D_{\mu} m_R | D_{\nu} n_R \rangle. \label{zeta_def}
\end{align}
Particularly, it can be seen that the sum in the relation for the NH-QGT is weighted by the overlap between left eigenvectors, giving a greater contribution to states that are closer to the state of interest. 

Note that every term in the sum (\ref{zeta_def}) is itself gauge invariant. 
This means that instead of summing over the entire spectrum we can limit ourselves to the term with $m=n$, defining a limited version of the NH-QGT
\begin{equation}\label{lim-NH-QGT}
    \Tilde{\zeta}_{\mu \nu}^{(n)} = \langle n_L | n_L \rangle \langle D_{\mu} n_R | D_{\nu} n_R \rangle.
\end{equation}
In principle, this quantity itself can be seen as a generalization of the QGT. It now depends only on one particular state and not the entire spectrum. Moreover, $\tilde \zeta_{\mu \nu}^{(n)}$ is Hermitian and positive-semidefinite in contrast to the full NH-QGT. Our calculations indicate that both $\zeta_{\mu \nu}$ and $\Tilde{\zeta}_{\mu \nu}$ can be used as a criticality measure. However, analytical results for $\zeta_{\mu \nu}$ are more feasible and transparent than those for $\Tilde{\zeta}_{\mu \nu}$, as will be seen below. 
Therefore, in the main text, we focus on the results for $\zeta_{\mu \nu}$, while a detailed discussion of $\Tilde{\zeta}_{\mu \nu}$ can be found in the SM.


In what follows we will illustrate that our NH-QGT introduced in this section is linked to the critical behavior of non-Hermitian systems.

\textit{\textbf{Non-Hermitian SSH model.}}
---
To demonstrate the capability of the NH-QGT in accurately detecting topological effects in non-Hermitian Hamiltonians, we apply it to the non-Hermitian analog of the Su-Schrieffer-Heeger (SSH) model \cite{Lieu_2018,Yin_2018}:
\begin{equation}
\begin{aligned}
    H =  (t+\delta) \sum_n a_n^{\dagger} b_n + (t-\delta) \sum_n b_n^{\dagger} a_n \\
    +t^{\prime} \sum_{n} (b_{n}^{\dagger} a_{n+1} + a^{\dagger}_{n+1} b_n),
\end{aligned}
\end{equation}
where $t^{\prime}$ represents the intercell hopping parameter, and the breaking of hermiticity arises from distinct intracell hopping parameters $t+\delta$ and $t-\delta$. The corresponding Bloch Hamiltonian for periodic boundary conditions is given by 
\begin{equation}\label{Bloch-Ham}
h(k) = \begin{pmatrix}
    0 && t-\delta + t^{\prime} e^{-i k} \\
    t+\delta + t^{\prime} e^{ik} && 0 
\end{pmatrix}
\end{equation}
and the energy levels are $\pm \sqrt{\varepsilon(k)}$, where $\varepsilon(k)= 1 + t^2 - \delta^2 + 2t \cos (k) - 2 i \delta \sin(k)$ (we put $t^{\prime} = 1$). 
Different phases are separated by the band crossing $\varepsilon(k)=0$. This occurs at $|t \pm \delta| = 1$. Thus, the parameter space is divided into four phases where $|t - \delta|$ and $|t + \delta|$ are either greater or less than one. In what follows we will denote these regions as $(s_1 , s_2)$ with $s_i = \pm$. Here $s_1 = +$ or $s_1 = -$ means that $|t-\delta|>1$ or $|t-\delta|<1$ correspondingly (similarly for $s_2$ and $|t+\delta|$).  
These phases are of topological nature as they can be characterized by two winding numbers of the Hermitian part of the Hamiltonian around exceptional points \cite{Yin_2018}.

Explicit expressions for $\eta_{\mu \nu}$~(\ref{LR-QGT}) in this model were obtained in~\cite{ye2023quantum}.
To calculate $\zeta_{\mu \nu}$ we can apply Eq.~(\ref{zeta_def}) to all $k$-sectors of the Hamiltonian independently. This leads to the NH-QGT of the following form:
\begin{equation}\label{zeta-NH-SSH}
    \begin{aligned}
        \zeta_{tt} = \sum_{k} \frac{\delta^2 + \sin^2(k)}{4 | \varepsilon|^2},& \quad \zeta_{\delta \delta} = \sum_{k} \frac{(t+ \cos(k))^2}{4 | \varepsilon |^2}, \\
        \zeta_{t\delta} = \zeta_{\delta t} =& -\sum_{k} \frac{ (t+\cos(k)) \delta }{4 |\varepsilon|^2},
    \end{aligned}
\end{equation}
where sums go over the Brillouin zone. It is already seen that singularities of this tensor are exactly at points where $\varepsilon(k)=0$, i.e. $|t \pm \delta| = 1$. It is even more apparent in the thermodynamic limit $L \rightarrow \infty$ where the NH-QGT becomes
\begin{equation}
\begin{aligned}
    &\zeta_{tt(\delta \delta)} = \frac{L}{16} \left( \frac{1}{| (\delta + t)^2 -1 |} + \frac{1}{| (\delta-t)^2-1 |} \pm f(\delta,t)\right), \\
    &\zeta_{t \delta} = \zeta_{\delta t} = \frac{L}{16} \left( \frac{1}{|(\delta+t)^2-1|} - \frac{1}{|(\delta-t)^2 - 1|} \right).
\end{aligned}
\end{equation}
Here the top (bottom) sign corresponds to the index $tt$ ($\delta \delta$) and the function $f(\delta, t)$ depends on the phase under consideration:
\begin{equation}
    f(\delta, t) = \begin{cases}
        \frac{2}{(\delta - t)(\delta + t)}, \quad & (+,+), \\
        \frac{1}{\delta(\delta + t)}, \quad & (- , +), \\
        \frac{1}{\delta(\delta - t)}, \quad & (+ , -), \\
        0, \quad & ( -, -).
    \end{cases}
\end{equation}
Each topological phase is described by its own tensor and the singularities of the NH-QGT are exactly at topological phase transition points. 
Note that $\zeta_{\mu \nu}$ has the Euclidean metric signature, in contrast to the Minkowski one for $\eta_{\mu \nu}$ in~\cite{ye2023quantum}.

\textit{\textbf{Dissipative quadratic models.}}
---
As our next testing ground, we use fermionic dissipative Markovian models whose time evolution is given by quadratic Liouvillians  $\partial_{t} \rho = \mathcal{L}(\rho)$ with
\begin{equation}\label{Liou}
    \mathcal{L}(\rho) = - i [ H , \rho] + \sum_{k} \left(  L_k \rho L_k^{\dagger} - \frac{1}{2} \{ L_k^{\dagger} L_k  , \rho \} \right), 
\end{equation}
where Hamiltonian $H = \boldsymbol{w}^T \boldsymbol{H} \boldsymbol{w}$ is quadratic and jump operators $L_k = \boldsymbol{l}_k^{T} \boldsymbol{w}$ are linear in Majorana operators $\boldsymbol{w} = 
(w_1 , ... \text{ }, w_{2n})^T$ that are defined as $w_{2j-1} = c_j + c_j^{\dagger}$, $w_{2j} = i (c_j - c_j^{\dagger})$. Here $c_j^{\dagger}$ and $c_j$ are the creation and annihilation operators of $n$ fermionic modes, $\boldsymbol{H}$ is a $2n \times 2n$ antisymmetric Hermitian matrix and $\boldsymbol{l}_k$ are arbitrary complex vectors of dimension $2n$.

Liouvilleans of this nature can be exactly diagonalized. One of the most elegant methods to achieve this is through the use of the so-called third quantization technique, which was introduced in \cite{Prosen_20088}. At the core of this method is a set of $2n$ superoperators that satisfy canonical anticommutation relations
\begin{equation}\label{third-quant}
    \hat{a}_j (\rho) = - \frac{i}{2} W [ w_j , \rho], \qquad \hat{a}^{\dagger}_j (\rho) = -\frac{i}{2} W \{ w_j , \rho \},
\end{equation}
where $W = i^n \prod_{l=1}^{2n} w_l$ and $\{ \hat{a}_j^{\dagger}, \hat{a}_k\} = \delta_{jk}$. Direct calculations show that the Liouvillean (\ref{Liou}) can be rewritten using these superoperators as
\begin{equation}\label{linb-third}
\mathcal{L} = - \sum_{ij} \left( X_{ij} \hat{a}_i^{\dagger} \hat{a}_i + \frac{1}{2} Y_{ij} \hat{a}_i^{\dagger} \hat{a}_j^{\dagger} \right),
\end{equation} 
where $\boldsymbol{X}= 4 i \boldsymbol{H} + 2\text{Re}\{ \boldsymbol{M}\} = \boldsymbol{X}^{*} $, $\boldsymbol{Y} = -4 i \text{Im} \{ \boldsymbol{M} \} = - \boldsymbol{Y}^{T} = \boldsymbol{Y}^{\dagger}$, and $\boldsymbol{M} = \sum_{k} \boldsymbol{l}_k \boldsymbol{l}_k^{\dagger} = \boldsymbol{M}^{\dagger} \geq 0$ is the so-called bath matrix. 

For simplicity we assume that the matrix $\boldsymbol{X}$ is diagonalizable $\boldsymbol{X} = U \boldsymbol{D}_X U^{-1}$ with $\boldsymbol{D}_X = \text{diag} ( \{ x_j \}_{j=1}^{2n} )$. In this case, Liouvillean can be diagonalized by a non-unitary Bogoluibov transformation~\cite{Blaizot1985QuantumTO} $\boldsymbol{b} = U^{-1} (\boldsymbol{a} + \Gamma \boldsymbol{a}^{\dagger})$, $\boldsymbol{b}^{\times} = U^{T} \boldsymbol{a}^{\dagger}$ and its diagonal form is given by $\mathcal{L}= - \sum_{j=1}^{2n} x_j b_j^{\times} b_j$. Here, the matrix $\Gamma$ is obtained from the solution of the Sylvester equation
\begin{equation}\label{Gamma-matrix}
    \boldsymbol{X} \Gamma + \Gamma \boldsymbol{X}^T = \boldsymbol{Y}.
\end{equation}
The uniqueness of the solution (which is equivalent to $\text{Min}_j \text{Re}(x_j) >0$) denotes the condition for the steady state to be unique. Furthermore, the NESS is a Gaussian state in terms of operators $w_j$ and the two-point correlation function of this state coincides with the matrix~$\Gamma$: $\langle [ w_i, w_j] \rangle_{\text{NESS}} = 2 \Gamma_{ij}$. 

The third quantization allows us to determine the AGP operator. The AGP with the required gauge (\ref{Gen-QGT-AGP}) for $\lambda$-dependent quadratic Liouvillean is also quadratic in fermions, and its structure resembles (\ref{linb-third}):
\begin{equation}\label{AGP-quadr}
    \mathcal{A}_{\lambda} = \sum_{ij} (\boldsymbol{{\cal X}}^{\lambda}_{ij} \hat{a}^{\dagger}_{i} \hat{a}_{j} + \frac{1}{2} \boldsymbol{{\cal Y}}^{\lambda}_{ij} \hat{a}^{\dagger}_i \hat{a}^{\dagger}_j ),
\end{equation}
where 
\begin{equation}\label{AGP-quad-XY}
    \boldsymbol{{\cal X}}^{\lambda} = U A_X^{(\lambda)} U^{-1}, \text{ } \boldsymbol{{\cal Y}}^{\lambda} =  \partial_{\lambda} \Gamma +   \boldsymbol{{\cal X}}^{\lambda} \Gamma + \Gamma (\boldsymbol{{\cal X}}^{\lambda})^T, 
\end{equation}
and $A_X^{(\lambda)}$ is the off-diagonal part of $\partial_{\lambda}U^{-1} U$. The detailed derivation of Eqs.~(\ref{AGP-quadr}) and~(\ref{AGP-quad-XY}) is given in the SM.  
Note that $\boldsymbol{{\cal X}}^{\lambda}$ is a real matrix, while $\boldsymbol{{\cal Y}}^{\lambda}$ is an imaginary anti-symmetric matrix. This structure of the AGP is imposed by its trace and hermiticity (but not positivity) preservation. 

Using the explicit form of the AGP~(\ref{AGP-quadr}) the NH-QGT in a steady-state can be calculated more explicitly:
\begin{equation}\label{NH-QGT-quadr}
    \zeta_{\mu \nu}^{\text{NESS}} = \frac{1}{2} \text{Tr} (\partial_{\mu} \Gamma \partial_{\nu} \Gamma) + \text{Tr} (\boldsymbol{{\cal X}}^{\mu} \Gamma \partial_{\nu} \Gamma).
\end{equation}
The first term in the relation~(\ref{NH-QGT-quadr}) encodes how dramatically the two-point correlation function changes in a steady-state. Given that Gaussian states are completely defined by their two-point correlation function, any non-analytical traits within the NESS essentially correspond to non-analyticity within the first term. The subsequent term contains both $\Gamma$ and $U$ matrices, illustrating the involvement of derivatives of other eigenstates in the generalized QGT [see Eq.(\ref{zeta_def})]. Here we note that in order to exclude the contribution of other eigenstates, one can consider the limited version $\Tilde{\zeta}_{\mu\nu}$ of the NH-QGT~(\ref{lim-NH-QGT}). However, in this case manifestation of the correlation matrix $\Gamma$ is more involved (see the SM).

 To illustrate this general framework with a specific example, we examine the Kitaev chain with local dissipation.  The Hamiltonian for this model, with periodic boundary conditions imposed, is given by  $\hat{H} = - \sum_{j=1}^L(c_j^{\dagger} c_{j+1} + \gamma c_j^{\dagger} c_{j+1}^{\dagger} + \frac{h}{2} c_j^{\dagger} c_j + \text{H.c.}) $. Additionally, two jump operators act on each site $L_{j}^{-} = g \mu_{-} c_j$, $L_{j}^{+} = g \mu_{+} c_j^{\dagger}$. This particular model was initially introduced in \cite{Cirac2013}. In the weak coupling limit $g \rightarrow 0 $,  the system becomes critical in the same regions as the Kitaev chain Hamiltonian: for $\gamma \neq 0$ there is a critical field $h=1$ and for $\gamma = 0$ the whole line $|h|<1$ is critical \cite{Cirac2013,Banchi_2014,carollo2017uhlmann}.

To compute the steady-state geometric tensor in this theory, we can utilize Eq.~(\ref{NH-QGT-quadr}). However, since the model exhibits translational invariance, it is beneficial to first apply the Fourier transform. Ultimately, the desired tensor is expressed as the sum over the Brillouin zone (detailed calculations can be found in the SM):
\begin{equation}
    \zeta_{\mu \nu}^{\text{NESS}} = \Lambda^2 \sum_{k} \sin^2(\varphi_k) \partial_{\mu} \varphi_k \partial_{\nu} \varphi_k, 
\end{equation}
where $\Lambda = (\mu^{2}_{+} - \mu^2_{-})/(\mu^2_{+} + \mu^2_{-})$, $\varphi_{k} = \text{arctg} \left(\frac{\gamma \sin (k)}{h - \cos(k)} \right)$, $\{ \lambda_1 , \lambda_2 \} = \{ h , \gamma \}$. The structure of the NH-QGT resembles the ground-state QGT for the Kitaev Hamiltonian \cite{Campos2007}, with the only difference being the additional term of $\sin^2(\varphi_k)$ under the summation. In the thermodynamic limit $L \rightarrow \infty$, one can replace the sums with integrals to derive analytical formulas. For instance, for $|h|<1$ the geometric tensor can be expressed as
\begin{equation}
    \zeta_{\mu \nu} = \frac{\Lambda^2 L }{8 |\gamma|} \text{diag}\left( \frac{3}{(1-h^2)} ,  \frac{1 + 3 |\gamma|}{(1+|\gamma|)^3} \right).
\end{equation}
Expressions for $|h|>1$ are given in the SM. This tensor exhibits singular behavior precisely at the critical regions.

\textit{\textbf{Concluding remarks.}}
--- Using the generator of adiabatic transformations we introduced a novel non-Hermitian generalization of the QGT.
We proposed that our generalization can serve as a universal tool for studying both dissipative and non-Hermitian phase transitions.
We examined our proposal using several exactly solvable models, including a non-Hermitian version of the Su-Schrieffer-Heeger (SSH) model and generic quasi-free fermionic Liouvillians. 
Furthermore, in the latter case we obtained an explicit expression for the AGP (super)operator.

Thus, our findings emphasize the significance of the adiabatic transformations in understanding phase transitions in non-Hermitian systems.
We expect our results to be valuable in the field of quantum control, particularly for investigating regimes of counter-diabatic drive in open quantum systems, similar to those in isolated systems~\cite{Kolodrubetz_2017}.
Another promising research direction is to explore potential connections between adiabaticity and dissipative quantum chaos, similar to those observed in Hermitian quantum systems~\cite{Pandey_2020, sá2023signatures}.

Finally, it should be emphasized that the Hermitian QGT is an experimentally measurable quantity. A possible method of measuring it in polaritonic systems was proposed in~\cite{Bleu_2018}.
This approach has recently been adapted for non-Hermitian polaritonic systems, thus providing a methodology for measuring~$\eta_{\mu \nu}$~\cite{Hu_2024}. We note that this method can also be straightforwardly utilized to measure the NH-QGT introduced in our work, enabling an experimental observation of this quantity.




\acknowledgements{
We thank V. Gritsev, T. Prosen, A. Rubtsov and his group for useful discussions. We are also grateful to A. Polkovnikov for useful comments on the manuscript. 
The research is supported by the Priority 2030 program at the National University of Science and Technology “MISIS” under the project K1-2022-027.
}

\bibliography{references}

\begin{thebibliography}{56}
\bibitem [{AGP()}]{AGPnote}%
  \BibitemOpen
  \href@noop {} {\bibinfo {title} {{Note that, unlike in the usual case, there is no imaginary unit in the definition of the AGP.}}}\BibitemShut {Stop}%
\bibitem [{Ide()}]{Identity-left-eigen}%
  \BibitemOpen
  \href@noop {} {\bibinfo {title} {{This is a consequence of the trace-preserving property of a Liouvillian. Indeed, $\text{Tr}(\mathcal{L}(\rho)) = 0$ for an arbitrary matrix $\rho$, so that $\mathcal{L}^{\dagger}(\hat{1}) = 0$}}}\BibitemShut {NoStop}%
\bibitem [{non()}]{non-diag}%
  \BibitemOpen
  \href@noop {} {\bibinfo {title} {{In the case of a non-diagonalizable matrix $K$ one needs to use the generalized eigenvectors of $K$ and then define the AGP via the response of these eigenvectors to a parameter change }}}\BibitemShut {NoStop}%
\end{thebibliography}

\clearpage

\begin{center}
\text{ \centering \LARGE Supplemental Material}
\end{center}

\section{Adiabatic Gauge Potential}

We consider a parameter-dependent non-Hermitian operator $K_{\lambda}$ with the right and the left eigenvectors
\begin{equation}
\begin{aligned}
    &K_{\lambda} \ket{n_R} = \Lambda_{n} \ket{n_R}, \\
    &K^{\dagger}_{\lambda} \ket{n_L} = \Lambda_{n}^* \ket{n_L}.
\end{aligned}
\end{equation}
These vectors form a biorthogonal basis $\langle n_L | m_R \rangle = \delta_{mn}$. To characterize their parameter-dependence one can introduce the generator of adiabatic transformations, known as the adiabatic gauge potential (AGP), as
\begin{equation}\label{AGP-right}
    \mathcal{A}_{\lambda} \ket{n_R} = \ket{\partial_{\lambda} n_R }.
\end{equation}
Then, differentiating the biorthogonality condition, one obtains
\begin{equation}\label{AGP-left}
    \mathcal{A}_{\lambda}^{\dagger} \ket{n_L} = - \ket{\partial_{\lambda} n_L}.
\end{equation}
Assuming that the operator $K$ is diagonalizable, we can write its spectral decomposition:
\begin{equation} \label{spectral_decomposition}
    K_{\lambda} = \sum_{n} \Lambda_{n}\ket{n_R} \bra{n_L}.
\end{equation}
Differentiating Eq.~(\ref{spectral_decomposition}) with respect to $\lambda$ and using Eqs.~(\ref{AGP-right}) and~(\ref{AGP-left}), we obtain
\begin{equation}\label{almost-equation}
    \partial_{\lambda} K_{\lambda} = \mathcal{F}_{\lambda} + [ \mathcal{A}_{\lambda} , K_{\lambda} ],
\end{equation}
where $\mathcal{F}_{\lambda} = \sum_{n} \partial_{\lambda} \Lambda_{n} \ket{n_R} \bra{n_L} $. As a result, one can derive the following operator equation for the AGP:
\begin{equation}\label{equation-for-AGP}
    [ \partial_{\lambda} K_{\lambda} - [ \mathcal{A}_{\lambda} , K_{\lambda} ]  , K_{\lambda} ] = 0.
\end{equation}
It follows immediately from Eq.~(\ref{equation-for-AGP}) that the AGP is defined up to an operator that commutes with $K_{\lambda}$. This can also be seen directly from the definition (\ref{AGP-right}). Indeed, the eigenvectors are defined up to a constant, so that we can always rescale them as
\begin{equation} \label{gauge_freedom_eigenstates}
    \ket{ n_R (\lambda) } \rightarrow e^{r_n (\lambda) }   \ket{ n_R (\lambda) },
\end{equation}
where $ r_n (\lambda) \in \mathbb{C}$ is an arbitrary complex-valued function. Then the AGP is transformed as 
\begin{equation}
 \mathcal{A}_{\lambda} \rightarrow \mathcal{A}_{\lambda} + {\cal R}_{\lambda}
\end{equation}
with ${\cal R}_{\lambda} = \text{diag}\{ r_n( \lambda ) \}$ is a diagonal matrix in the basis $\ket{n_R}$. Using the gauge freedom~(\ref{gauge_freedom_eigenstates}) we can always put $\langle n_L | \mathcal{A}_{\lambda} | n_R \rangle $ to zero for all $n$. Another consequence that can be seen from Eq.~(\ref{almost-equation}) is that the off-diagonal matrix elements ($m \neq n$) of the AGP in an instantaneous basis read
\begin{equation}\label{A-matrix}
    \langle m_L(\lambda) | \mathcal{A}_{\lambda} | n_R(\lambda) \rangle = \frac{  \langle m_L|\partial_{\lambda} K | n_R \rangle }{ \Lambda_{n} - \Lambda_{m}}.
\end{equation}
Clearly, the matrix element~(\ref{A-matrix}) is ill-defined in the case of degeneracy. To resolve this issue, one can define a regularized version of the AGP operator as
\begin{equation}\label{A-matrix-reg}
   \langle m_L(\lambda) | \mathcal{A}_{\lambda} (\mu) | n_R(\lambda) \rangle =  \frac{\Lambda_{n}^{*} - \Lambda_{m}^{*}}{|\Lambda_{n} - \Lambda_{m}|^2 + \mu^2}  \langle m_L|\partial_{\lambda} K | n_R \rangle, 
\end{equation}
where $\mu$ is an energy cut-off. 
The regularized AGP in Eq.~(\ref{A-matrix-reg}) is useful for numerical evaluation of the NH-QGT~(\ref{Gen-QGT-AGP}).

\section{Dissipative Quadratic Models and AGP}

\subsection{General case}
Using the notations introduced in the main text [see Eqs.~(\ref{Liou}) and~(\ref{third-quant})], we start by writing the Liouvillian in a third quantized form:
\begin{equation}
    \mathcal{L} = - \sum_{r,s} \left( X_{r,s} \hat{a}^{\dagger}_r \hat{a}_s + \frac{1}{2} Y_{r,s} \hat{a}^{\dagger}_r \hat{a}^{\dagger}_s  \right).
\end{equation}
In order to diagonalize this Liouvillian superoperator it is useful to rewrite it in the matrix form (we follow Ref.~\cite{Banchi_2014} for the diagonalization procedure)
\begin{equation}
    \mathcal{L} = - \frac{1}{2} 
    \begin{pmatrix}
    \boldsymbol{a}^{\dagger} & \boldsymbol{a}
    \end{pmatrix}
    \boldsymbol{L}
    \begin{pmatrix}
        \boldsymbol{a} \\
        \boldsymbol{a}^{\dagger}
    \end{pmatrix} - \frac{1}{2} \text{tr} \boldsymbol{X},
\end{equation}
where we introduced the block matrix
\begin{equation}
    \boldsymbol{L}= 
    \begin{pmatrix}
        \boldsymbol{X} & \boldsymbol{Y} \\
        0 & -\boldsymbol{X}^T
    \end{pmatrix}.
\end{equation}
Firstly, we observe that the matrix $\boldsymbol{L}$ can be reduced to a block diagonal form by the transformation
\begin{equation}
    \boldsymbol{T} = 
    \begin{pmatrix}
        \boldsymbol{1} & \Gamma \\
        \boldsymbol{0} & \boldsymbol{1}
    \end{pmatrix},
    \quad \boldsymbol{T}^{-1} = 
    \begin{pmatrix}
        \boldsymbol{1} & - \Gamma \\
        \boldsymbol{0} & \boldsymbol{1}
    \end{pmatrix},
\end{equation}
where the $2n\times 2n$ matrix $\Gamma$ satisfies Eq.~(\ref{Gamma-matrix}). Indeed, in this case one can easily obtain
\begin{equation}
    \boldsymbol{T} \boldsymbol{L}\boldsymbol{T}^{-1} = 
    \begin{pmatrix}
        \boldsymbol{X} & \boldsymbol{0} \\
        \boldsymbol{0} & -\boldsymbol{X}^T
    \end{pmatrix}.
\end{equation}
In the next step one diagonalizes the matrix $\boldsymbol{X}$:
\begin{equation}
    U^{-1} \boldsymbol{X} U = D_X = \text{diag}(\{x_k\}_{k=1}^{2n}).
\end{equation}
Thus, one can introduce the transformation
\begin{equation}
    \boldsymbol{S} = 
    \begin{pmatrix}
        U^{-1} & 0\\
        0 & U^T
    \end{pmatrix} \boldsymbol{T},
\end{equation}
which diagonalizes the Liouvillian matrix $\boldsymbol{L}$ as
\begin{equation}
    \boldsymbol{L} = S^{-1} \boldsymbol{D} S \quad \text{with} \quad  \boldsymbol{D} = \begin{pmatrix}
        D_{X} & \boldsymbol{0} \\
        \boldsymbol{0} & - D_X.
    \end{pmatrix}
\end{equation}
This allows us to rewrite the Liouvillian in the diagonal form
\begin{equation}
 \mathcal{L} = - \frac{1}{2}  \begin{pmatrix}
    \boldsymbol{b}^{\times} & \boldsymbol{b}
    \end{pmatrix}
     \boldsymbol{D}
    \begin{pmatrix}
        \boldsymbol{b} \\
        \boldsymbol{b}^{\times}
    \end{pmatrix} - \frac{1}{2} \text{tr} \boldsymbol{X},
\end{equation}
where 
\begin{equation}
    \begin{pmatrix}
        \boldsymbol{b} \\
        \boldsymbol{b}^{\times}
    \end{pmatrix} = S
    \begin{pmatrix}
        \boldsymbol{a} \\
        \boldsymbol{a}^{\dagger}
    \end{pmatrix}
\end{equation}
is a non-unitary Bogoluibov transformation~\cite{Carollo_2020}.

Now, we proceed with finding the AGP operator for quasi-free dissipative fermionic models. Using the fact that the two operators of the form
\begin{equation}
\begin{aligned}
    \mathcal{A} = \frac{1}{2} \begin{pmatrix}
    \boldsymbol{a}^{\dagger} & \boldsymbol{a}
    \end{pmatrix}
    \boldsymbol{A}
    \begin{pmatrix}
        \boldsymbol{a} \\
        \boldsymbol{a}^{\dagger}
    \end{pmatrix} , \quad \mathcal{B} = \frac{1}{2} \begin{pmatrix}
    \boldsymbol{a}^{\dagger} & \boldsymbol{a}
    \end{pmatrix}
    \boldsymbol{B}
    \begin{pmatrix}
        \boldsymbol{a} \\
        \boldsymbol{a}^{\dagger}
    \end{pmatrix}
\end{aligned}
\end{equation}
commute as
\begin{equation}
    [ \mathcal{A} , \mathcal{B}  ] = \frac{1}{2} \begin{pmatrix}
    \boldsymbol{a}^{\dagger} & \boldsymbol{a}
    \end{pmatrix}
     [ \boldsymbol{A} , \boldsymbol{B} ]
    \begin{pmatrix}
        \boldsymbol{a} \\
        \boldsymbol{a}^{\dagger}
    \end{pmatrix},
\end{equation}
we can solve Eq.~(\ref{equation-for-AGP}) for the AGP operator explicitly, which yields
\begin{equation}
    \mathcal{A}_{\lambda} = \frac{1}{2}  \begin{pmatrix}
    \boldsymbol{b}^{\times} & \boldsymbol{b}
    \end{pmatrix}
     \partial_{\lambda} S S^{-1}
    \begin{pmatrix}
        \boldsymbol{b} \\
        \boldsymbol{b}^{\times}
    \end{pmatrix}
\end{equation}
where
\begin{equation}
    \partial_{\lambda} S S^{-1} = \begin{pmatrix}
        \partial_{\lambda} U^{-1} U & U^{-1} \partial_{\lambda} \Gamma (U^{-1})^T \\
        \boldsymbol{0} & - (\partial_{\lambda} U^{-1} U)^T
    \end{pmatrix}.
\end{equation}
Also, removing the diagonal part of the last matrix, we find the AGP operator with the removed diagonal matrix elements: 
\begin{equation}\label{beautiful}
    \mathcal{A}_{\lambda} =  \frac{1}{2} 
    \begin{pmatrix}
    \boldsymbol{a}^{\dagger} & \boldsymbol{a}
    \end{pmatrix}
    \begin{pmatrix}
        \boldsymbol{{\cal X}}^{\lambda} & \boldsymbol{{\cal Y}}^{\lambda} \\
        0 & -(\boldsymbol{{\cal X}}^{\lambda})^T
    \end{pmatrix}
    \begin{pmatrix}
        \boldsymbol{a} \\
        \boldsymbol{a}^{\dagger}
    \end{pmatrix},
\end{equation}
where
\begin{equation}\label{AGP-quad-2}
    \boldsymbol{{\cal X}}^{\lambda} = U A_X^{(\lambda)} U^{-1}, \text{ } \boldsymbol{{\cal Y}}^{\lambda} =  \partial_{\lambda} \Gamma +   \boldsymbol{{\cal X}}^{\lambda} \Gamma + \Gamma (\boldsymbol{{\cal X}}^{\lambda})^T 
\end{equation}
and $A_X^{(\lambda)}$ is the off-diagonal part of $\partial_{\lambda}U^{-1} U$. After that, it is easy to obtain the second NH-QGT. Rewriting the AGP operator and its conjugated version in terms of the $\boldsymbol{b}$ modes and acting on the steady-state (which is a $\boldsymbol{b}$-vacuum), one finds
\begin{equation}\label{genQGT-quadr}
    \zeta_{\mu \nu}^{\text{NESS}} = \frac{1}{2} \text{Tr} (\partial_{\mu} \Gamma \partial_{\nu} \Gamma) + \text{Tr} (\boldsymbol{{\cal X}}^{\mu} \Gamma \partial_{\nu} \Gamma).
\end{equation}

\subsection{Translationally invariant case}
In the presence of the translational symmetry we relabel the Majorana fermions as
\begin{equation}
    \boldsymbol{w}_j = \begin{pmatrix}
        w_{j,1} , w_{j,2} 
    \end{pmatrix}^T
\end{equation}
where $j$ is a spatial index. Then the Hamiltonian and the bath matrix take the following form:
\begin{equation}
\begin{aligned}
    &\mathcal{H} = \sum_{j,r} \boldsymbol{w}_j^{T} h (j-r) \boldsymbol{w}_r, \\
    & \boldsymbol{M}_{(j,\beta)(r,\beta')} = m(j-r)_{\beta \beta'}. 
\end{aligned}
\end{equation}
Thus, the Liouvillian is given by
\begin{equation}
    \mathcal{L} = - \frac{1}{2} \sum_{j,r} 
    \begin{pmatrix}
    \boldsymbol{a}^{\dagger}_j & \boldsymbol{a}_j
    \end{pmatrix}
    \boldsymbol{L}(j-r)
    \begin{pmatrix}
        \boldsymbol{a}_r \\
        \boldsymbol{a}^{\dagger}_r
    \end{pmatrix} - \frac{1}{2} \text{tr} \boldsymbol{X}
\end{equation}
and we can apply the Fourier transform 
\begin{equation}
    \begin{pmatrix}
        a_j \\
        a_j^{\dagger}
    \end{pmatrix} =  \frac{1}{\sqrt{N}} 
    \sum_{k} e^{i k j} 
    \begin{pmatrix}
        a_k \\
        a_{-k}^{\dagger}
    \end{pmatrix}
\end{equation}
in order to obtain
\begin{equation}
    \mathcal{L} = - \frac{1}{2} \sum_{k} 
    \begin{pmatrix}
    \boldsymbol{a}^{\dagger}_k & \boldsymbol{a}_{-k}
    \end{pmatrix}
    \boldsymbol{L}(k)
    \begin{pmatrix}
        \boldsymbol{a}_k \\
        \boldsymbol{a}^{\dagger}_{-k}
    \end{pmatrix} - \frac{1}{2} \text{tr} \boldsymbol{X},
\end{equation}
where
\begin{equation}
\begin{aligned}
    &\boldsymbol{L} (k) = \sum_{r} e^{- i k r} \boldsymbol{L} (r)
 = 
    \begin{pmatrix}
        \boldsymbol{x}(k) & \boldsymbol{y} (k) \\
        0 & -\boldsymbol{x}^{T}(-k)
    \end{pmatrix}, \\
    &\boldsymbol{x} (k) = 4 i \boldsymbol{h}(k)+ m(k) + m^{T} (-k), \\
    &\boldsymbol{y} (k) = -2 ( m(k) - m^{T} (-k) ).
\end{aligned}
\end{equation}
After that, the Liouvillian can be diagonalized in each $k$-sector independently using the machinery from the previous subsection. In this case the AGP operator becomes
\begin{equation}
    \mathcal{A}_{\lambda} = \frac{1}{2} \sum_{k} \begin{pmatrix}
    \boldsymbol{a}^{\dagger}_k & \boldsymbol{a}_{-k}
    \end{pmatrix}
    \begin{pmatrix}
        \boldsymbol{{\cal X}}^{\lambda} (k) & \boldsymbol{{\cal Y}}^{\lambda} (k) \\
        0 & -(\boldsymbol{{\cal X}}^{\lambda}(-k))^T
    \end{pmatrix}
    \begin{pmatrix}
        \boldsymbol{a}_k \\
        \boldsymbol{a}^{\dagger}_{-k}
    \end{pmatrix}, 
\end{equation}
where 
\begin{equation}
\begin{aligned}
    &\boldsymbol{{\cal X}}^{\lambda}(k) = u(k) A_x^{(\lambda)}(k) u^{-1}(k), \\&\boldsymbol{{\cal Y}}^{\lambda} (k) =  \partial_{\lambda} \gamma (k) +   \boldsymbol{{\cal X}}^{\lambda}(k) \gamma(k) + \gamma(k) (\boldsymbol{{\cal X}}^{\lambda}(-k))^T,
\end{aligned}
\end{equation}
and $A_x^{(\lambda)}(k)$ is the off-diagonal part of $\partial_{\lambda}u^{-1}(k) u(k)$. Here $\gamma(k)$ is a matrix that solves the equation
\begin{equation}\label{gamma-fourier-eq}
    \boldsymbol{x}(k) \gamma (k) + \gamma(k) \boldsymbol{x}^T (-k) = \boldsymbol{y}(k),
\end{equation}
and $u(k)$ diagonalizes the matrix $\boldsymbol{x}(k)$
\begin{equation}
    u^{-1} x(k) u = \begin{pmatrix}
        x_1 (k) & 0 \\
        0 & x_2 (k)
    \end{pmatrix}.
\end{equation}
The expression for the second NH-QGT is similar to that in Eq.~(\ref{genQGT-quadr}), but with the summation carried over the Brillouin zone:
\begin{equation}\label{QGT-fourier}
    \zeta_{\mu \nu} = \sum_k \left( \frac{1}{2}\text{Tr}(\partial_{\mu} \gamma \partial_{\nu} \gamma ) +  \text{Tr}( \boldsymbol{{\cal X}}^{\lambda}(k) \gamma \partial_{\nu} \gamma ) \right).
\end{equation}

\section{NH-QGT in the Kitaev chain with local dissipation}
The Hamiltonian of the Kitaev chain can be written in terms of Majorana fermions as
\begin{multline}
    \mathcal{H} = - \sum_{j} \left( \frac{1 + \gamma}{2} i w_{j,2} w_{j+1,1} + \frac{1-\gamma}{2} (-i) w_{j,1} w_{j+1,2} \right) \\
    -i h \sum_{j} w_{j,1} w_{j,2} = \sum_{r,s} \boldsymbol{w}_r^{T} h (r-s) \boldsymbol{w}_s.
\end{multline}
Therefore, the non-zero matrices $h(j)$ are given by
\begin{equation}
\begin{aligned}
    &h(0) = 
    \begin{pmatrix}
        0 & -\frac{i h }{2} \\
        \frac{i h }{2} & 0
    \end{pmatrix}, \\
    &h(1) =
    \begin{pmatrix}
        0 & i \frac{1 + \gamma}{4} \\
        -i \frac{1-\gamma}{4} & 0
    \end{pmatrix}, \\
    &h(-1) = h^{\dagger}(1) = 
    \begin{pmatrix}
        0 & i \frac{1-\gamma}{4} \\
        -i \frac{1 +\gamma}{4}
    \end{pmatrix}.
\end{aligned}
\end{equation}
The Fourier transform of the Hamiltonian is
\begin{equation}
    h(k) = \frac{1}{2} \gamma \sin (k) \sigma_x + \frac{1}{2}(h-\cos(k)) \sigma_y.
\end{equation}
The dissipation is described by the jump operators acting on each lattice site
\begin{equation}
\begin{aligned}
   L_j^{+} = g \mu_{+} c_j^{\dagger} = \frac{g \mu_{+}}{2} ( w_{j,1} + i w_{j,2}), \\
   L_j^{-} = g \mu_{-} c_{j} = \frac{g \mu_- }{2} (w_{j,1} - i w_{j,2}),
\end{aligned}
\end{equation}
leading to a bath matrix of the form
\begin{equation}
    m(k) = \frac{g^2 (\mu^2_+ + \mu^2_-)}{4} \boldsymbol{\sigma}_0 + \frac{g^2 (\mu^2_+ - \mu^2_-)}{4} \boldsymbol{\sigma}_y.
\end{equation}
Finally, for this model we find
\begin{equation}
\begin{aligned}
    &\boldsymbol{x} (k) = \frac{g^2}{2}(\mu^2_+ + \mu^2_-) \boldsymbol{\sigma}_0 + 2 i \gamma \sin (k) \sigma_x + 2 i (h-\cos(k)) \boldsymbol{\sigma}_y, \\
    &\boldsymbol{y}(k) = - g^2 (\mu^2_+ - \mu^2_- ) \boldsymbol{\sigma}_y.
\end{aligned}
\end{equation}
Using these expressions, one can find $\boldsymbol{\gamma} (k)$ from Eq.~(\ref{gamma-fourier-eq}). In the weak coupling limit $g \rightarrow 0$ it is given by 
\begin{equation}
    \boldsymbol{\gamma} (k) = - \Lambda \cos (\varphi_k) \sin (\varphi_k) \boldsymbol{\sigma}_x - \Lambda \cos^2 (\varphi_k) \boldsymbol{\sigma}_y,
\end{equation}
where we have introduced $\Lambda = \frac{\mu^{2}_+ - \mu^2_-}{\mu^2_+ + \mu^2_-}$ and $\varphi_k = \text{arctg} \left( \frac{\gamma \sin (k)}{h - \cos (k)} \right)$. The matrix $\boldsymbol{x} (k)$ can be diagonalized by the following transformation
\begin{equation}
    u(k) = \frac{1}{\sqrt{2}} 
    \begin{pmatrix}
        e^{i\varphi_k/2} & i e^{i \varphi_k/2} \\
        i e^{-i\varphi_k/2} & e^{- i\varphi_k/2}
    \end{pmatrix}.
\end{equation}
Hence, the NH-QGT in the steady-state (\ref{QGT-fourier}) reads
\begin{equation}
    \zeta_{\mu \nu}^{\text{NESS}} =  \Lambda^2 \sum_{k} \sin (\varphi_k^2) \partial_{\mu} \varphi_k \partial_{\nu} \varphi_k.
\end{equation}
In the thermodynamic limit $L \rightarrow \infty$ all sums are replaced by the integrals $\sum_k \rightarrow \frac{L}{2 \pi} \int dk$ and we obtain

\begin{widetext}

\begin{equation}
    \zeta_{hh} = \frac{3}{8}  \Lambda^2 L \times
    \begin{cases}
        \frac{1}{|\gamma| (1-h^2)}, \quad |h|<1 \\
        \frac{\gamma^4 | h |}{(h^2-1) (h^2 + \gamma^2 -1)^{5/2}}, \quad |h| > 1 
    \end{cases}
\end{equation}

\begin{align}
    &\zeta_{\gamma \gamma} (|h|<1) = \Lambda^2 L \frac{1+3 |\gamma|}{8 |\gamma| (1 + |\gamma|)^3},\\
    \zeta_{\gamma \gamma}(|h|>1) = \Lambda^2 L \Biggl( \frac{ \gamma^2}{(1-\gamma^2)^3} +  \frac{3}{8} & \frac{\gamma^2}{(h^2 + \gamma^2 -1)^{5/2}} - \frac{5}{4} \frac{|h| \gamma^2}{(1-\gamma^2)^3 (h^2 + \gamma^2 -1)^{1/2} } + \frac{1}{8} \frac{|h| \gamma^2 ( 2 h^4 - 5 (1-\gamma^2)^2)}{(1-\gamma^2)^3 (h^2 + \gamma^2 -1)^{5/2}}\Biggr)
\end{align}

\begin{equation}
    \zeta_{\gamma h} = \frac{3}{8} \Lambda^2 L \times
    \begin{cases}
        0, \quad |h|<1 \\
        - \frac{\text{sign}(h) \gamma^3}{ (h^2 + \gamma^2 -1)^{5/2}}, \quad |h|>1.
    \end{cases}
\end{equation}
\end{widetext}

\section{Calculation of $\Tilde{\zeta}_{\mu \nu}$}
In this section we discuss in more detail the tensor
\begin{equation}\label{zeta-tilde}
    \Tilde{\zeta}_{\mu \nu}^{(n)} = \langle n_L | n_L \rangle  \langle D_{\mu} n_R | D_{\nu} n_R \rangle,
\end{equation}
which was introduced in Eq.~(\ref{lim-NH-QGT}) of the main text. Clearly, $\Tilde{\zeta}_{\mu \nu}^{(n)}$ is invariant under the gauge transformations of eigenstates. Indeed, the covariant derivatives make the second factor in Eq.~(\ref{zeta-tilde}) to transform up to a constant [see Eq.~(\ref{cov_derivatives_transform}) in the main text], whereas the first factor simply removes this constant, guaranteeing the gauge invariance. At the same time, $\Tilde{\zeta}_{\mu \nu}^{(n)}$ is Hermitian, positive-semidefinite, and it reduces to the conventional QGT in the Hermitian case. This motivates us to calculate this tensor in the models discussed in the main text in order to compare the results with those obtained with the NH-QGT $\zeta_{\mu \nu}^{(n)}$.

For Liouvillian steady-states the corresponding left eigenvector can be always fixed to be the identity operator. Then, the steady-state simply becomes a usual normalized density matrix and a covariant derivative can be replaced by an ordinary one. Hence, for a steady-state, from Eq.~(\ref{zeta-tilde}) one obtains
\begin{equation}\label{zet-til-dens}
    \Tilde{\zeta}_{\mu \nu}^{\text{NESS}} = {\cal D} \text{Tr} (\partial_{\mu} \rho_{\text{ss}} \partial_{\nu} \rho_{\text{ss}}),
\end{equation}
where ${\cal D}$ is the dimension of the corresponding Hilbert space and $\rho_{\text{ss}}$ is the NESS density matrix. Clearly, for NESSes $\Tilde{\zeta}_{\mu \nu}^{\text{NESS}}$ is real and symmetric. Also note that when the steady-state is a pure state, Eq.~(\ref{zet-til-dens}) reduces (up to a numerical factor) to the Fubini-Study metric, i.e., the symmetric part of the Hermitian QGT.

It is instructive to compare this quantity with a mixed state generalization of the QGT -- the Quantum Fisher Tensor~\cite{Ercolessi_2012,CarolloGEOMofQPT}. The Fisher Tensor is introduced using a logarithmic derivative $G_{\mu}$ of a density matrix defined via the relation $\partial_{\mu} \rho_{\text{ss}}  = G_{\mu} \rho_{\text{ss}} + \rho_{\text{ss}} G_{\mu}$.
Then, the Fisher Tensor is an average of a product of log-derivatives with respect to the density matrix $Q_{\mu \nu} \equiv \text{Tr} ( \rho_{\text{ss}} G_{\mu} G_{\nu})$, while its symmetric part, known as the Bures metric, is $g^{B}_{\mu \nu} = \frac{1}{2} \text{Tr}(\rho_{\text{ss}} \{ G_{\mu} , G_{\nu}\})$. Using the log-derivatives, Eq.(\ref{zet-til-dens}) can be rewritten in the form
\begin{equation}
    \Tilde{\zeta}_{\mu \nu}^{\text{NESS}}  = {\cal D}  \text{Tr} (\rho^2_{\text{ss}} \{G_{\mu} , G_{\nu} \}) + 2 {\cal D}  \text{Tr}( \rho_{\text{ss}} G_{\mu} \rho_{\text{ss}} G_{\nu} ).
\end{equation}
This formula suggests to rescale our tensor as 
\begin{equation}\label{rescaled-tensor}
\Tilde{\zeta}_{\mu \nu}^{(n)} \rightarrow \frac{1}{\langle n_L | n_L \rangle \langle n_R | n_R \rangle} \Tilde{\zeta}_{\mu \nu}^{(n)}  = \frac{\langle D_{\mu} n_R | D_{\nu} n_R \rangle}{\langle n_R | n_R \rangle}
\end{equation}
since after that its value for the NESS can be interpreted as an average over $\rho_1 = \frac{1}{\text{Tr}(\rho_{\text{ss}}^2)}\rho_{\text{ss}}^2 $, which is a well defined density matrix itself:
\begin{equation}\label{tilde-NESS}
    \Tilde{\zeta}_{\mu \nu}^{\text{NESS}}  =  \text{Tr} (\rho_1 \{G_{\mu} , G_{\nu} \}) + 2  \text{Tr}( \rho_1^{1/2} G_{\mu} \rho_1^{1/2} G_{\nu} ).
\end{equation}
In this light it is clear that the $\Tilde{\zeta}_{\mu \nu}$ goes in parallel with the definition of the Bures metric, although its geometrical interpretation is somewhat blurry.

Recasting $\Tilde{\zeta}_{\mu \nu} $ using the log-derivatives is also useful from the computational point of view. For Gaussian Fermionic states an explicit expression for the log-derivative was obtained in Ref.~\cite{carollo2017uhlmann}. Its form is given by
\begin{equation}
    G_{\mu} = \frac{1}{4} \boldsymbol{w}^T \boldsymbol{K}_{\mu} \boldsymbol{w} + \frac{1}{4} \text{Tr}(\boldsymbol{K}_{\mu} \Gamma),
\end{equation}
where $\Gamma_{ij} = \frac{1}{2} \text{Tr}(\rho_{\text{ss}} [w_i , w_j])$ is the correlation matrix in a Gaussian state and $\boldsymbol{K}_{\mu}$ is defined via the equation
$\Gamma \boldsymbol{K}_{\mu} \Gamma - \boldsymbol{K}_{\mu} = \partial_{\mu} \Gamma$. For a Gaussian NESS $\rho_1$ is also a Gaussian state, but with the correlation matrix $\Gamma_1 = \frac{2\Gamma}{1+\Gamma^2}$. All this allows us to express solely in terms of correlation matrix both the Bures metric (that was firstly done in \cite{Banchi_2014})
\begin{equation}
\begin{aligned}
    g_{\mu \nu}^{B} = \frac{1}{8} \text{Tr} \left( \partial_{\mu} \Gamma  \frac{1}{1-\text{Ad}_{\Gamma}}(\partial_{\nu} \Gamma) \right)   \\
    =\frac{1}{8} \sum_{j,k} \frac{(\partial_{\mu}\Gamma)_{jk} (\partial_{\nu}\Gamma)_{kj}}{1-\gamma_j \gamma_k}
\end{aligned}
\end{equation}
and the tensor we are interested in (\ref{tilde-NESS})
\begin{equation}
\begin{aligned}
    \Tilde{\zeta}_{\mu \nu}^{\text{NESS}} = \frac{1}{2} \text{Tr} \left( \frac{1}{1 + \Gamma^2} \partial_{\mu} \Gamma \frac{1}{1+\Gamma^2} \partial_{\nu} \Gamma \right)  \\
    = \frac{1}{2} \sum_{j,k} \frac{(\partial_{\mu}\Gamma)_{jk} (\partial_{\nu}\Gamma)_{kj}}{(1+\gamma_j^2)(1+\gamma_k^2)}.
\end{aligned}
\end{equation}
Here $\text{Ad}_{\Gamma}(X) = \Gamma X \Gamma$ and both summations are performed in the basis where $\Gamma$ is diagonal, with the eigenvalues being $\gamma_j$. Using this closed form expression the same steps as for the Bures metric~\cite{Banchi_2014} can be applied to elucidate the connection of $\Tilde{\zeta}_{\mu \nu}^{\text{NESS}}$ to criticality. For instance, in the Kitaev chain described above the explicit form of this tensor is given by
\begin{equation}
    \Tilde{\zeta}_{\mu \nu}^{\text{NESS}} = \Lambda^2 \sum_{k} \frac{\partial_{\mu} \varphi_k \partial_{\nu} \varphi_k}{\left(1+\Lambda^2 \text{cos}^2(\varphi_k)\right)^2}.
\end{equation}
Due to an obvious inequality $ 1\leq 1+\Lambda^2 \text{cos}^2(\varphi_k) \leq 1+\Lambda^2$ it can be seen that singular regions of $\Tilde{\zeta}_{\mu \nu}^{\text{NESS}}$ are exactly the same as for the Hermitian QGT, i.e., it is singular on the critical lines.

For the case of the NH-SSH the situation is even more transparent. The corresponding left and right eigenstates for the Bloch Hamiltonian (\ref{Bloch-Ham}) are given by
\begin{equation}
\begin{aligned}
\ket{\psi_R^{(\pm)} (k)} = \frac{1}{\sqrt{2}} \begin{pmatrix}
    \pm 1 && \frac{t+\delta+e^{i k}}{\varepsilon^{1/2} }
\end{pmatrix}^T \\
\bra{\psi_{L}^{(\pm)}(k)} = \frac{1}{\sqrt{2}} \begin{pmatrix}
    \pm 1 && \frac{t-\delta+e^{-i k}}{ \varepsilon^{1/2}}
\end{pmatrix}
\end{aligned}
\end{equation}
with $\varepsilon(k)= 1 + t^2 - \delta^2 + 2t \cos (k) - 2 i \delta \sin(k)$. Explicit form of the eigenstates allows one to calculate any quantity of interest directly. In this particular model it turns out that the rescaled $\Tilde{\zeta}_{\mu \nu}$ (\ref{rescaled-tensor}) coincides exactly with the second NH-QGT~(\ref{zeta-NH-SSH}).

Finally, let us make a brief comment on the physical meaning of both $\zeta_{\mu \nu}$ and $\Tilde{\zeta}_{\mu \nu}$. Summing $\zeta_{\mu \nu}$ over the whole spectrum, one obtains the norm of the AGP operator, an important characteristic of the system that describes the magnitude of change of the entire spectrum~\cite{Pandey_2020}. Thus, finding the AGP norm requires calculation of $\zeta_{\mu \nu}$ over the whole spectrum. However, if one tries to compare the magnitude of the change in, say, two specific states from the spectrum, it seems that it is more natural to characterize this not by the matrix elements of $\mathcal{A}^{\dagger}_{\mu} \mathcal{A}_{\nu}$ (which is $\zeta_{\mu \nu}$), but describing how projectors $P_n = \ket{n_R} \bra{n_L}$ on these eigenstates are deformed. Specifically, for a change in some parameter $\lambda$ one has
\begin{multline}
    \frac{|| \partial_{\lambda} P_n ||^2}{|| P_n ||^2} = \frac{\langle D_{\lambda} n_R | D_{\lambda} n_R \rangle}{\langle n_R |  n_R \rangle} + \frac{\langle D_{\lambda} n_L | D_{\lambda} n_L \rangle}{\langle n_L |  n_L \rangle} \\
    + \frac{\langle  n_L | D_{\lambda} n_L \rangle \langle  n_R | D_{\lambda} n_R \rangle }{\langle n_L |  n_L \rangle \langle n_R |  n_R \rangle} + \frac{\langle  D_{\lambda} n_L |  n_L \rangle \langle  D_{\lambda} n_R |  n_R \rangle }{\langle n_L |  n_L \rangle \langle n_R |  n_R \rangle}.
\end{multline}
As one can see, this quantity contains the rescaled tensor~$\Tilde{\zeta}_{\mu \nu}$ from Eq.~(\ref{rescaled-tensor}) as well as its left-state counterpart. We will elaborate more on this connection in the future work related to adiabaticity and dissipative chaos.

\end{document}